\documentclass[runningheads]{llncs}

\usepackage{booktabs} 
\usepackage{hyperref}
\usepackage{multirow}
\usepackage{float}
\usepackage{xcolor}
\usepackage{dingbat}
\usepackage{graphicx}

\usepackage[linesnumbered]{algorithm2e}

\begin{document}
\title{EXmatcher: Combining Features Based on Reference Strings and Segments to Enhance Citation Matching}
\titlerunning{EXmatcher}

\author{Behnam Ghavimi \and
Wolfgang Otto  \and
Philipp Mayr}

\institute{GESIS -- Leibniz Institute for the Social Sciences\\
\email{firstname.lastname@gesis.org\\}}

\authorrunning{B. Ghavimi et al.}

\maketitle 

\begin{abstract}
Citation matching is a challenging task due to different problems such as the variety of citation styles, mistakes in reference strings and the quality of identified reference segments.
The classic citation matching configuration used in this paper is the combination of blocking technique and a binary classifier. Three different possible inputs (reference strings, reference segments and a combination of reference strings and segments) were tested to find the most efficient strategy for citation matching. 
In the classification step, we describe the effect which the probabilities of reference segments can have in citation matching. 
Our evaluation on a manually curated gold standard showed that the input data consisting of the combination of reference segments and reference strings lead to the best result. In addition, the usage of the probabilities of the segmentation slightly improves the result.
\end{abstract}


\section{Introduction}
The need for data integration from various sources is growing in information systems in order to improve data quality and reusability of the data e.g. for retrieval or data analysis.
The procedure of finding records in a database that correspond to the same entity (e.g. files, publications, data sets, ...) across another data set is typically called record linkage. Record linkage has been used in different domains~\cite{christen2012data,herzog2007data}.
The application of record linkage in the domain of bibliographic data is known as citation matching or reference matching.
High quality citation data for research publications is the basis for areas like bibliometrics but also for integrated digital libraries (DL). Citation data are valuable since they show the linkage between publications. The extraction of reference information from full text is called citation extraction. One key challenge for the aforementioned tasks is to match the extracted reference information to given DLs.
The process of mapping an extracted reference string to one entity of a given DL is called citation matching \cite{lawrence1999digital}.
Proper citation matching is an essential step for every citation analysis \cite{moed2006citation} and the improvement of citation matching leads to a higher quality of bibliometric studies. In the DL context, citation data is one important source of effective information retrieval, recommendation systems and knowledge discovery processes \cite{mayr-IJDL2018}.

Despite the widely acknowledged benefits of citation data, the open access to them is still insufficient.
Some commercial companies such as Clarivate Analytics, Elsevier or Google possess citation data in large-scale and use them to provide services for their users. 
Recently, some initiatives and projects e.g. the "Open Citations" project or the "Initiative for Open Citations" focus on publishing citation data openly\footnote{\url{https://i4oc.org/}}. The "Extraction of Citations from PDF Documents" -   EXCITE\footnote{\url{http://excite.west.uni-koblenz.de/website/}} project is one of these projects. The aim of EXCITE is extracting and matching citations from social science publications \cite{koerner2017} and making more citation data available to researchers. With respect to this objective, a set of algorithms for information extraction and matching has been developed focusing on social science publications in the German language. The shortage of citation data for the international and German social sciences is well known to researchers in the field and has itself often been subject to academic studies \cite{moed2006citation}.  

This paper is dedicated to the step of citation matching in the EXCITE pipeline and the responsible algorithm for this task is called EXmatcher\footnote{\url{https://github.com/exciteproject/EXmatcher}}. For the matching task in EXCITE, different target databases/DLs are defined a) sowiport \cite{Hienert2015}, b) GESIS Search\footnote{\url{https://search.gesis.org/}} and c) Crossref\footnote{\url{https://search.crossref.org}}. 
The matching target for the study in this paper is solely sowiport.
Sowiport contains bibliographic metadata records of more than 9 million references on publications and research projects in the social sciences.\\
This paper makes the following contributions: 
\begin{itemize}
    \item Introduction of a gold standard for the citation matching task, 
    \item Evaluate the effect of different inputs in the citation matching steps and
    \item Investigate the effect of the utilization of reference segmentation probabilities as features in the citation matching procedure.
\end{itemize} 
The remainder of this paper is structured as follows. In section 2, we organize the related work around the concepts of the record linkage pipeline known from~\cite{christen2012data}.
Section 3 describes the set-up of citation matching in the EXCITE project.
Section 4 is about creating a citation matching gold standard corpus and evaluation of our algorithm with different configurations.
Finally, section 5 summarizes the key outcomes of our improvements on citation matching.

\section{Related Work}
Christen et al.~\cite{christen2012data} suggested general steps for the matching process after reviewing different matching approaches: \textbf{(1)} Input pre-processing, \textbf{(2)} Blocking technique, \textbf{(3)} Feature extraction and classification. EXmatcher also follows these steps for citation matching and considers different input configurations to investigate their affects. In the following, we organize the related work according to these steps. 

\subsection{Input pre-processing}
As the first step, input data need to be pre-processed in a way that it becomes proper for the matching algorithm. 
To identify similar strings during all parts of the matching process a common method to increase robustness is to normalize the input strings. A simple normalization is to lowercase the input string and remove punctuation and stop words.

If an algorithm depends on reference segments for matching, this data need to be extracted from the reference strings. 
PDFX \cite{constantin2013pdfx},
Exparser~\cite{jcdlyezd}, 
GROBID \cite{lopez2009grobid}, ParsCit \cite{councill2008parscit} are few examples of tools that perform reference segmentation. Wellner et al.~\cite{wellner2004integrated} investigated the effect of extraction probabilities on citation matching by the consideration of different number of best Viterbi segmentations. EXmatcher considers only the best Viterbi segmentation and uses the probability of each segment in feature vector provided for a binary classifier regarding the citation matching task.   

Phonetic function is another technique used in this step and the common idea behind all phonetic encoding functions is that they attempt to convert a string into a code based on pronunciation~\cite{christen2006comparison}. Phonetic algorithms are mainly used for name segments. Pre-processing functions can also be used in other steps. For example, data which has been prepared by phonetic functions can be considered as blocking keys in the indexing step since indexing brings similar values together. These techniques can also be used in the feature extraction step to generate vectors of features for classifiers. This encoding process is often language dependent. 
Soundex algorithm was developed by Russell and Odell in 1918 ~\cite{odell1918soundex} for English language pronunciation. 
Phonex~\cite{lait1996assessment}, NYSIIS~\cite{taft1970name}, and Cologne functions~\cite{postel1969kolner} are some other examples of phonetic functions.
The Cologne phonetics is based on the Soundex phonetic algorithm and is optimized to match the German language. We also used Cologne phonetic function in our implementation since our main focus was working on German language papers.

\subsection{Blocking Technique}
The next step is the blocking technique in order to decrease the number of pairs required to be compared. Imagine, we need to match a set of $n$ references extracted from publications to a bibliographic database with $m$ entries. In a naive way, comparisons of every reference with every entry in the database are required which results in a complexity of $n\times m$. Considering a set of 100,000 references and a database with 10 million bibliographic entries this results in $10^{12}$ comparisons. 
In the blocking approach, we split target and source into blocks of data depending on a common attribute or combination of the attributes.
After finding corresponding blocks in source and target we reduce the number of necessary comparisons to the number of combinations between corresponding blocks.
For example, we have a reference with "2001" as publication year, so in this case, it is not necessary to compare this reference with the entire records in the target database. We only could compare to the entries in the block of records published in 2001.
In related works (e.g.,~\cite{olensky2016evaluation}), even they use blocks related to one year before and after that year. 

Several blocking or indexing techniques have been introduced till now~\cite{christen2012data}.
As an example, D-Dupe~\cite{kang2008interactive} is a tool implemented for data matching and network visualizations. D-Dupe implemented an indexing technique based on standard blocking~\cite{fellegi1969theory}. 
Hernandez et al. suggested a sorted neighborhood approach~\cite{hernandez1995merge,hernandez1998real}.
This technique, instead of generating a key for each block, sorts the data for matching based on a 'sorting key'. In suffix or q-gram based indexing approaches there is a higher chance to have correct matches in a same block since the idea behind them are for handling different forms of entities and errors. 
In the citation matching field, Fedoryszak et al. presented a blocking method based on hash functions~\cite{fedoryszak2014efficient}.
Another research field deals with the identification of efficient blocking keys.
Koo et al. tried to find the best combination of citation record fields~\cite{koo2011effects} that helps increase citation matching performance.

\subsection{Feature Extraction and Classification}
In the third step of the citation matching process, each candidate record pair (i.e, reference (string and related segments) and each retrieved item by blocking) are compared using a variety of attributes and comparison functions. The output of this step is a feature vector for each pair. In the final step, each compared candidate record pair is classified into one of the classes (i.e., match, non-match) using the related feature vector.

Comparison functions such as Jaro-Winkler, Jaccard, or Levenshtein are often used for analyzing textual values.
As an example, the D-Dupe tool includes string comparison functions such as Levenshtein distance, Jaro, Jaccard, and Monge-Elkan~\cite{kang2008interactive}. 

For the classification step of citation matching the reference can be represented by a reference string or by extracted segments. Also a combination of both is possible as we show in this work.

Foufoulas et al.~\cite{foufoulas2017high} suggested an algorithm which matches reference strings without reference segmentation. Their approach first tries to detect the reference section by some heuristic and then attempts to identify the title of a record in the target repository in the reference section. Finally, it validates this match with more metadata of the record in the target repository. Their title detection and citation validation steps are mostly based on the combination of simple search and comparison functions. One of classification approaches is a threshold-based one.
In this type, the similarity between vectors of two items will be calculated (e.g., by using cosine similarity algorithm) and if the similarity score is higher than a predefined threshold, then two items are matched. 

A rule-based classification employs some rules for classification~\cite{cohen2000data,hernandez1998real,naumann2010introduction}. These rules consist of a combination of smaller parts and the link between these parts are logical "AND", "OR" and "NOT" operands. These rules define the similarity of pairs. In the optimal case, each rule in a set of rules should have a high precision and recall~\cite{mining2006data}. 
More strict or specific rules usually have high precision, while general or easy rules often have low precision but high recall.
The iterative, rule-based citation matching algorithm of CWTS (Center for Science and Technology Studies)~\cite{olensky2016evaluation} relies on a series of matching rules. These rules are applied iterative in decreasing order of strictness. The citation matching algorithm starts with the most restrictive matching rules (e.g., exact match on first author, publication year, publication title, volume number, starting page number, and DOI). Afterward, it proceeds with less restrictive matching rules (e.g. match on Soundex encoding of the last name of the first author, publication year plus or minus one, volume number, and starting page number). The less restrictive matching rules allow for various types of inaccuracies in the bibliographic fields of cited references. In all rules, the Levenshtein distance is used to match the publication name of a cited reference to the publication name of a cited article.

Viewing probabilistic record linkage from a Bayesian perspective has also been discussed by Fortini et al.~\cite{fortini2001bayesian} and Herzog et al.~\cite{herzog2007data}. If training data are available, then a supervised classification approach can be employed. Many binary classification techniques have been introduced~\cite{mining2006data,mitchell1997artificial}, and many of these techniques are used for matching. Decision tree is one of these supervised classification techniques~\cite{mining2006data}. 
As an example, Cochinwala et al.~\cite{cochinwala2001efficient} build a training set and trained a Regression Tree (CART) classifier~\cite{Breiman1984Classification} for data matching.
TAILOR tool~\cite{elfeky2002tailor} for data matching uses e.g. a ID3 decision tree. 

The Support Vector Machine (SVM) classification algorithm~\cite{Vapnik2000} is based on the idea of mapping the input data of the classifier into a higher dimensional vector space using a kernel function.
This is done to be able to separate samples for the target classes using a hyperplane even if it is not possible in the lower dimension.
SVM as a large margin classifier optimizes during training time through maximization of the distance between training samples and hyperplane. 
Fedoryszak et al.~\cite{fedoryszak2013large} presented a citation matching solution using Apache Hadoop. Their algorithm is based on reference segments and also uses SVM algorithm to confirm the status of items (i.e., match or not match) based on the created features.

\section{Matching Procedure in EXCITE}
\subsection{Input Data for Matching}
\label{sssec:DPMEXcite}
For matching we used two types of information from each reference, the raw reference strings and structured information (i.e. segments).
The segmentation is done with Exparser\footnote{\url{https://github.com/exciteproject/Exparser}} which is a CRF-based algorithm \cite{jcdlyezd}.
The output of Exparser includes a probability for each predicted segment.
This information is taken into account as an additional information to enhance the results of the matching procedure. 
To enhance the results for the publication year information we extract year mentions from the raw reference strings with a regular expression independently from the parser.
We also remove extra characters in the year segment (e.g. b in 1989b).
As a last pre-processing step we have combined volume and issue to one segment called number because during parsing the issue was often recognized as a volume and vice versa.

\subsection{Blocking Step}
\label{sssec:bsu}

We used the search platform Solr\footnote{\url{http://lucene.apache.org/solr/}} for blocking.
For each reference, EXmatcher retrieves the corresponding block with the help of blocking queries.
The whole blocking procedure is described in Algorithm 1.
\begin{algorithm}
 \KwData{Pre-processed reference $r$, indexed bibliographic database $D$, cutoff parameter $c$}
 \KwResult{A set of suggested matching records $S$}
 \BlankLine
 Generate a query set $Q$ based on segments or reference string\;
 Initialize an empty suggestion set $S$\;
 \ForEach{query q in query set $Q$}{
  Retrieve ranked result list with query $q$\ as $Rl$\;
  \If{size of result list $Rl$ $\geq$ 0}{
    Cut off ranked result list at position $c$\;
    Join reduced list $Rl$ to $S$\;
   }
 }
 \caption{Blocking step for matching in EXmatcher}
\end{algorithm}
First, queries are formulated with the help of the parsed segments and the raw reference strings.
Therefore we used the operators OR and AND from the Solr query syntax\footnote{\url{https://lucene.apache.org/solr/guide/6\_6/query-syntax-and-parsing.html}}.
Additional we use fuzzy search ($\sim$-operator) which reflects a fuzzy string similarity search based on the Levenshtein distance.

The output of the blocking step is a ranked lists of retrieved items from the target database.
The items are ranked by the Lucene score based on tf/idf\footnote{https://lucene.apache.org/core/7\_0\_0/core/index.html?org/apache/lucene/search/\\similarities/TFIDFSimilarity.html}. 
To get the best trade off between retrieving all possible matching items and the reduction of necessary comparisons in the following classification task we identified two opportunities for influence.
One is varying the query and select the best query formulation. The other is the selection of a cut off threshold which determines how many of the retrieved items per query are used for further processing. 

As it is mentioned, firstly, queries out of segment combinations should be generated. For six segments (i.e., 1-Author, 2-Title, 3-Year, 4-Page, 5-Number (Volume/Issue), 6-Source) this results in a maximum of 63 segment combinations. For each query generated based on one of combinations needs to have one correct information about each of the segments queried. For example, if year of publication and authors' names are used, one of the author names has to be correct and also the year of publication. For title and source we used a fuzzy query on the whole segment string. For numbers at least one found number have to be in the volume or the issue field of the record in our database. To exclude not well performing segment combinations for query generation we measure the precision at one of the queries on our gold data. We only select segment combinations where at least 60\% of the retrieved items are a correct match. This reduces the number of maximum combinations we consider for query generation by 25\% to 48.

As an alternative strategy we generated queries only from the reference strings without using information from the segmentation. This strategy tries to deal with the problem that title information is often not correctly identified during segmentation.
But since the title is the most effective field for matching, the following approach is used which can act independently from the quality of the segmentation. For this we consider all token of the reference string as potentially including title information. The idea is to formulate a bigram search of the whole reference string. The resulting query leads to results which at least need to include one bigram of the reference string in the title field. But the more bigrams of the reference string are included in the title, the more preferred results are. Therefore a query based on only these bigrams of the reference string will be added to the set of queries. In addition, to increase the precision, a query based on year and bigrams of the reference string will also be considered. For this the year information is taken into account which is extracted with a regular expression.

The effect on blocking for the two strategies of query generation and even a comparison with a mixture of both strategies is described in~\ref{ssec:evaluation-blocking}.

\subsection{Classification for citation matching task}
\label{sssec:cfcmt}
After retrieving candidates for matches with our blocking procedure we need to decide which of the found candidate our system identify as a match.
I.e. the decision if the retrieved item is a match and hence the reference and the entry in the database are representing the same entity.  

For this we train and evaluate a binary classifier which is able to judge a pair of reference and match candidate as match or non match.
It is worth noting, that our approach is able to handle duplicates in our reference database.

The crucial step for building this classifier is feature selection.
We combine features generated from the raw reference string and from the segmentation.
One novelty of our approach is to test the usefulness of utilizing the certainty of our parser for the detected segments as an additional input feature for our classifier.
The output of Exparser contains for each token of all segments a probability value reflecting the certainty of the model.
If we have a high probability for a segment, the chance of having a wrong predicted label is low. 
Therefore, we expect that the usage of features reflecting this probabilities will have a noticeable effect on the performance of citation matching.

The first group of features is based on the comparison of the reference segments and the retrieved items in the blocking step:
\begin{itemize}
\item Some example of features based on the author segment:
    \begin{enumerate}
        \item Levenshtein score (phono-code and exact),
        \item Segmentation probability of first author (surname) 
    \end{enumerate}
\item Some example of features based on titles and source:
    \begin{enumerate}
        \item Jaccard score (including segmentation probabilities),
        \item Levenshtein score (token and letter level)
    \end{enumerate}
\item Some example of features based on numbers, pages, and publication year:
    \begin{enumerate}
        \item Jaccard score, and 
        \item Segmentation probability
    \end{enumerate}
\end{itemize}
An example for the usage of the probability is the extended version of the Jaccard score for author names.
The Jaccard similarity for the last names is the intersection of last names over the union of the set of last names in two records.
If the size of the intersection of the last names of two records is 2 and the size of the union of them is 4, then the Jaccard score would be 0.5 ((1+1)/4).
Our enhanced metric uses the extracted probabilities as weights in the intersection. If probabilities of these items in the intersection of lasts names are 0.8 and 0.9, then the new Jaccard score would be 0.42 ((0.8+0.9)/4). For the creation of the features of the second group all information based on segmentation is excluded.
These are features based only on the comparison of the raw reference string with the information of the retrieved record.
You can find some examples of this group in the following list:
\begin{enumerate}
    \item Longest common sub-string of title and reference string, and
    \item Occurrence of the abbreviation of the source field (e.g., journal abbreviation) in index in reference string.
\end{enumerate}

\section{Evaluation}
\label{SS:Evaluation}
\subsection{Gold Standard for Matching Algorithm}
\label{SS:gsfmE}
The computation of off-line evaluation metrics such as precision, recall and F-measure need ground truths. 
A manually checked gold standard was generated to assess the performance of the algorithms.
For creating this gold standard, we applied a simple matching algorithm based on blocking on a randomly selected set of reference strings from the EXCITE corpus. 
The EXCITE corpus contains SSOAR\footnote{\url{https://www.gesis.org/ssoar/home/}} corpus (about 35k), SOJ:Springer Online Journal Archives 1860-2001 corpus (about 80k), and Sowiport papers (about 116K). We used queries based on different combinations of title, author and publication year segments and considered the top hit in the retrieved blocks based on the Solr score.
The result was a document id (from sowiport) for each reference, if the approach could find any match. In the second step, these ids detected by the matching algorithm were completed by duplication information in sowiport to reach a list of all candidates of match items for each references. 

Afterwards, a trained human assessor checked the results.
If the previous step leads to an empty result set the assessor was asked to retrieve a record based on manually curated search queries. These manual queries used the information from the correct segments by manually extracting them from the reference strings.
If the corresponding item was found, it was added to the gold standard.
It also appeared, that not only one match was found, but also duplicates. 
In this case the duplicates where also added as matching items.
When matching items are found in the previous step, the assessor checked this list to remove wrong items and add missing items.
The result of this process is a corpus containing 816 reference strings. 517 of these items have at least one matched item in sowiport.
We published this corpus and a part of sowiport data (18,590 bibliographic items) openly for interested researchers in our Github repository\footnote{\url{https://github.com/exciteproject/EXgoldstandard/tree/master/Goldstandard\_EXmatcher}}.
\subsection{Evaluation of Blocking Step}
\label{ssec:evaluation-blocking}
\begin{figure}[h!]
\centering
\begin{minipage}{.5\textwidth}
  \includegraphics[width=\textwidth]{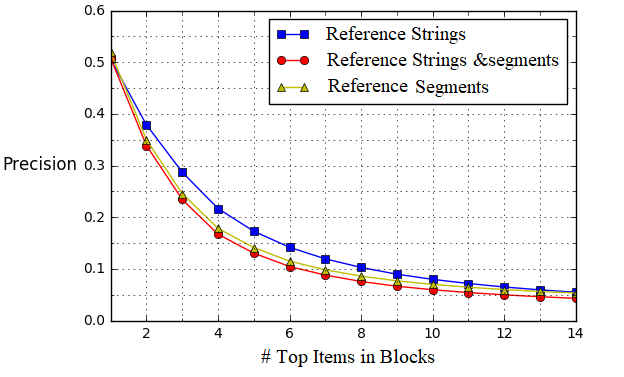}
    \caption{Precision of blocking} 
    \label{fig:precisionExmatchdup}
\end{minipage}%
\begin{minipage}{.465\textwidth}
  \centering
    \includegraphics[width=\textwidth]{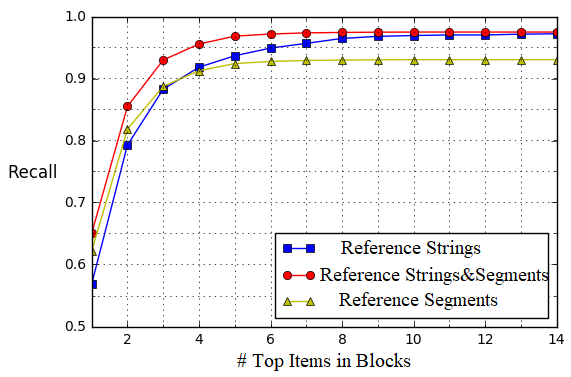}
    \caption{Recall of blocking} 
    \label{fig:recallExmatchdup}
\end{minipage}
\end{figure}

In this evaluation, three different configurations for input of blocking (i.e., 1- using \textit{only reference strings}, 2- using \textit{only reference segments}, and 3- the \textit{combination of reference segments and strings}) were examined. In addition, the effect of the consideration of different numbers of top items from the blocking step was checked. Fig.~\ref{fig:precisionExmatchdup} shows that the precision curve of blocking based on reference strings is higher than the two other configurations. This is not a big surprise because using only reference strings in our approach means focusing on the title and year fields (which it is explained in section~\ref{sssec:DPMEXcite} and section~\ref{sssec:bsu}) and the usage of these two fields has a high precision score to retrieve items.  

On the one hand by considering more items of the blocking list the precision is decreasing.
On the other hand the recall shown in Fig.~\ref{fig:recallExmatchdup} reach a score higher than 0.9 after the consideration of the 4 top items of blocking.
The highest recall has been achieved using the combination of reference strings and segments. Surprisingly, the curve of reference strings become closer to the combination of reference string and segments by consideration of more top items in blocking and almost reach to that in number 14. All these three curves become almost steady after consideration of 11 top retrieved items for each blocking query. 

Since we have another step after blocking which improve the precision, the important point in blocking is keep recall score high and at the same time shrinking the number of items for comparison. The precision of these three curves were not significantly different, therefore, the combination of reference strings and segments is picked in blocking step to generate input for the evaluation of classification step. 
For the number of top items in blocking, which are used for further processing in our pipeline, five is selected because considering more then five items is not leading to a higher recall value.

The selected configuration leads to a number of 1 to 39 retrieved items per reference.
The average number was 14 records with a standard deviations of 6.5.

For the 816 references of the gold standard 10,997 match candidates are generated with our configuration.
For each pair of reference and corresponding match candidate in our reference database sowiport we know if it is a match or not based on our gold standard.
In these 10,997 pairs, 1,026 (9.3\%) are correct matches and 9,971 (90.7\%) are no matches.
After blocking, the number of reference strings which have at least one correct match is 507, and 302 references are without any correct pair.
It means only ten references (1.2\%) which have at least one match in the gold standard could not pass blocking successfully, i.e. blocking step could not suggest any correct match for them.

\subsection{Evaluation of Classification Step}
\begin{figure}
\centering
\begin{minipage}[t]{.47\textwidth}
      \includegraphics[width=\textwidth]{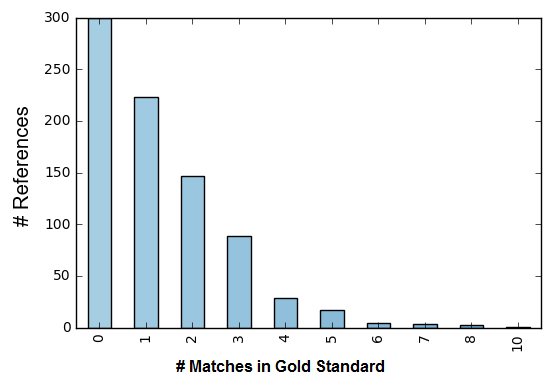}
    \caption{Frequency of references in gold standard with the number of matches in target database sowiport}
    \label{fig:numberofgoldstandard}
\end{minipage}%
\hfill
\begin{minipage}[t]{.47\textwidth}
  \centering
    \includegraphics[width=\textwidth]{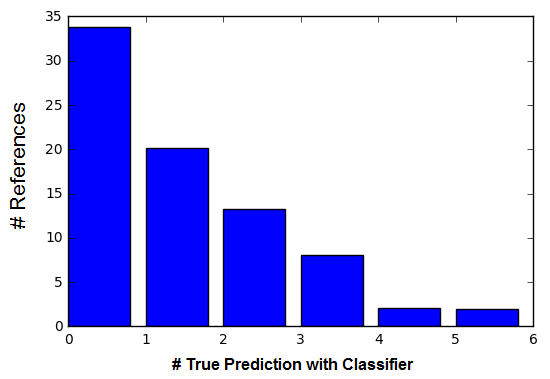}
    \caption{Average number of references in folds with true prediction of match class}
    \label{fig:duppre}
\end{minipage}
\end{figure}

In this section we present the results of the classification task.
We applied ten-fold cross validation for testing different classifier and feature combinations.
Blocking generated results for 809 references were split into ten separated groups and their related pairs placed in the related group to form the ten folds for cross validation.
Table~\ref{classeval} contains precision, recall and f-measure for our compared configurations.
\begin{table*}[h!]
\centering
\begin{tabular}{|l|l|l|l|l|l|l|l|}
\hline
Ref\_String & Ref\_Segments & Seg\_probability & SVM   & Random Forest    & Precision & Recall & F1    \\ \hline
\checkmark       & \checkmark         & \checkmark             & \checkmark  & - & 0.947 *     & 0.904  & 0.925 * \\ \hline
\checkmark        & \checkmark          & \checkmark             & - & \checkmark  & 0.938     & 0.906  & 0.921 \\ \hline \hline
\checkmark        & \checkmark          & -            & \checkmark  & - & 0.941     & 0.908 *  & 0.924 \\ \hline
\checkmark        & \checkmark         & -            & -& \checkmark & 0.923     & 0.899  & 0.910 \\ \hline \hline \hline
-       & \checkmark          & \checkmark             & \checkmark  & - & 0.942     & 0.865  & 0.901 \\ \hline
-      & \checkmark          & \checkmark             & -  & \checkmark  & 0.918     & 0.874  & 0.895 \\ \hline \hline 
-      & \checkmark         & -            & \checkmark  & - & 0.836     & 0.869  & 0.852 \\ \hline
-       & \checkmark          & -            & - & \checkmark  & 0.876     & 0.883  & 0.879 \\ \hline \hline \hline
\checkmark       & -         & -           & \checkmark & -  & 0.843     & 0.903  & 0.871 \\ \hline
\checkmark       & -         & -           & - & \checkmark  & 0.879     & 0.855  & 0.866 \\ \hline
\end{tabular}
\caption{Evaluation macro-metrics of different classifiers including duplicate matches for each reference string - the highest value in each column is marked by * symbol.}
\label{classeval}
\end{table*}
The results show that the SVM classifier using the combination of reference strings and segment features with considering segmentation probability has the highest F1 and precision scores. The second highest F1 score is related to the SVM classifier which uses the combination input features but this time without the segmentation probabilities.
The interpretation of data is that using the combination of inputs (reference segments and strings) has the main impact on the accuracy scores. 
The average number of references in different folds based on their number of correct predictions of the classifier with the highest F1 score are shown in Fig.~\ref{fig:duppre}.

In most real world scenarios it is only necessary to find exactly one match in a bibliographic database.
Because of this we evaluate our matching algorithm again.
For this evaluation only one correct match have to be found for each reference.
Regarding this purpose, we pick the highest probability generated by the classifier among match pairs for each reference \footnote{The threshold for decision between two classes would be the 0.5 - default threshold for the SVM classifier in scikit-learn python package.}. 
For this evaluation, we used the combination of features based on reference strings and segments as the input (including segments probabilities). In this case, average precision and recall scores for SVM algorithm are 0.97 and 0.92. For random forest algorithm, average precision and recall scores are 0.96 and 0.93.
To calculate a final score for the complete pipeline, also 10 references which could not pass blocking step should be considered.
Since the consideration of these items changes the number of false negatives, we see the effect on the recall score. Consequently, recall in the pipeline with SVM would be 0.913 and the pipeline using the Random Forest classifier would be 0.917. These evaluation scores are included in Table~\ref{my-labeSvm67}.

\begin{table}[h!]
\centering
\caption{10-fold cross-validation of SVM classifier regarding finding only one match for each reference}
\label{my-labeSvm67}
\begin{tabular}{|l|c|c|c|c|c|c|}
\hline
    & Precision & Recall & F1 & Precision-Pipeline & Recall-Pipeline & F1-Pipeline \\ \hline
SVM & 0.972 & 0.926 & 0.948 & 0.972 & 0.913 & 0.941 \\ \hline
Random & 0.967 & 0.931  & 0.948 & 0.967 & 0.917 & 0.941 \\ \hline
\end{tabular}
\end{table}

\section{Discussion and Conclusions}
In this paper, we explained our approach for handling the task of citation matching in the EXCITE project.
The implemented algorithm (EXmatcher) follows the classic solution for this task which contains three steps (i.e., 1- data normalization and cleaning, 2- blocking, 3- feature vector creation and classification). 
We analyzed the impact of different inputs (i.e., reference strings, segments and the combination of both) on the performance of our citation matching algorithm.
In addition, we investigated the benefit of using segments probabilities in the citation matching task.
The segmentation probabilities are considered directly and as weights for creating specific features for the classifier of EXmatcher. 

Using the combination of reference strings and segments as input with a SVM classification outperforms the other configurations in terms of F1 and precision scores.
Segments probabilities have a good impact on the precision score when the citation matching algorithm uses segments as input. 
For example, in the configuration of using only segments as the input and using SVM, segments probabilities can improve the precision about 11\% (Table~\ref{classeval}).
The combination of reference strings and segments can also cover the effect of considering segments probabilities. It means including/excluding segments probability doesn't affect the accuracy when citation matching algorithm uses the combination of two input data.
The effect of utilizing different classifiers on the result are very depended on other parameters in the citation matching configuration such as input types (i.e. reference strings, segments or both) and the consideration of segment probabilities.
The combination of reference strings and segments as the input for citation matching shows a higher recall than using each of them alone. 
But still 10 references which have at least one match couldn't pass the blocking step with using the combination the both data. One reason of this incident was that in generating queries,  EXmatcher combines the information from reference strings and information from reference segments in one query and links them with OR logical operand. Decreasing the number of failed in blocking step leads to a higher recall. One solution could be to send queries based on reference strings and based on segments in different queries and then the algorithm combines the retrieved items. 
Also more items can be extracted from reference strings input (such as pages, issue, volume and DOI) with some rule based steps and used in the blocking. 

The citation matching approach which has been described and evaluated in this paper is implemented in a demonstrator which connects all important steps from reference extraction, reference segmentation and matching in the EXCITE toolchain (see \cite{hosseini2019} \url{http://excite.west.uni-koblenz.de/excite}).

\bibliographystyle{splncs04}


\bibliography{bibliography}

\end{document}